\documentclass[prl,twocolumn,superscriptaddress,nobibnotes,a4paper, nofootnoteinbib]{revtex4-2}
\pdfoutput=1

\usepackage{graphicx}
\usepackage{physics}
\usepackage[dvipsnames]{xcolor}
\usepackage[separate-uncertainty = true,multi-part-units=single]{siunitx}
\usepackage{dcolumn}
\usepackage{bm}
\usepackage[version=4]{mhchem}
\usepackage[colorlinks=true,linkcolor=blue,urlcolor=blue,citecolor=blue]{hyperref}
\usepackage{natbib}
\usepackage{ulem}

\def\Var{{\textrm{Var}}\,}

\usepackage{mhchem}
\usepackage{siunitx}

\begin{document}

\title{Optical readout of a superconducting qubit using a piezo-optomechanical transducer}\thanks{This work was published in \href{https://doi.org/10.1038/s41567-024-02742-3}{Nat.\ Phys.\ \textbf{21}, 401--405 (2025).}}

\author{T.C.~van~Thiel}
\author{M.J.~Weaver}
\author{F.~Berto}
\author{P.~Duivestein}
\author{M.~Lemang}
\author{K.L.~Schuurman}
\author{M.~\v{Z}emli\v{c}ka}
\author{F.~Hijazi}
\author{A.C.~Bernasconi}
\author{C.~Ferrer}
\author{E.~Cataldo}
\affiliation{QphoX B.V., Elektronicaweg 10, 2628XG, Delft, The Netherlands}
\author{E.~Lachman}
\author{M.~Field}
\author{Y.~Mohan}
\affiliation{Rigetti Computing Inc., 775 Heinz Avenue, Berkeley, California, 94710, United States}%
\author{F.K.~de~Vries}
\author{C.C.~Bultink}
\author{J.C.~van Oven}
\affiliation{Qblox B.V., Delftechpark 22, 2628XH, Delft, The Netherlands}
\author{J.Y.~Mutus}
\affiliation{Rigetti Computing Inc., 775 Heinz Avenue, Berkeley, California, 94710, United States}%
\author{R.~Stockill}\email{rob@qphox.eu}
\author{S.~Gr\"oblacher}\email{simon@qphox.eu}
\affiliation{QphoX B.V., Elektronicaweg 10, 2628XG, Delft, The Netherlands}

\begin{abstract}
Superconducting quantum processors have made significant progress in size and computing potential. However, the practical cryogenic limitations of operating large numbers of superconducting qubits are becoming a bottleneck for further scaling. Due to the low thermal conductivity and the dense optical multiplexing capacity of telecommunications fiber, converting qubit signal processing to the optical domain using microwave-to-optics transduction would significantly relax the strain on cryogenic space and thermal budgets. Here, we demonstrate optical readout of a superconducting transmon qubit through an optical fiber connected via a coaxial cable to a fully integrated piezo-optomechanical transducer. Using a demolition readout technique, we achieve a single shot readout fidelity of $81\%$. Our results illustrate the benefits of piezo-optomechanical transduction for low-dissipation operation of large quantum processors.
\end{abstract}

\maketitle

Quantum computers are rapidly evolving from experimental settings towards commercially developed systems with ever increasing numbers of qubits~\cite{NatureEditorial2022}. Several important milestones have recently been achieved, including the first computations of problems that are practically infeasible to solve with even the largest classical computers~\cite{arute2019quantum, zhong2020quantum}. Nevertheless, solving tasks that are not merely proof-of-principle in character and could be of commercial interest, in particular with fault tolerant operation, are still far beyond the reach of even the most advanced quantum processing units (QPUs)~\cite{Brooks_2023}. Superconducting processors, one of the leading architectures for QPUs, currently operate using a few hundred qubits, with roadmaps for scaling up to thousands~\cite{kjaergaard2020superconducting, chow2021ibm, quantumai2023, rigetti2023}. Scaling QPUs to these comparably small sizes will already require significant improvements in cryogenic input-output technologies and necessitate dilution refrigerators which are many times larger than current systems~\cite{bluefors2023}. With existing technology, much of the capacity in physical space and cooling power will be consumed by the amplifiers, circulators and coaxial cables required for controlling and reading out qubits. Reaching the millions of qubits anticipated to be required for fault-tolerant quantum computing applications~\cite{Preskill2018quantumcomputingin} therefore remains an elusive goal without significant technological leaps. Mitigating these limitations by converting microwave signals to the optical domain is a promising approach to tackle this challenge and has attracted significant interest~\cite{mirhosseini2020superconducting, han2020cavity, han2021microwave, wang2022high, meesala2023non, jiang2023optically}.

Transducing qubit state-information to the optical domain allows for a reduced passive heat load of the read-out chain by up to three orders of magnitude~\cite{lecocq2021control}, as the thermal conductivity of optical fibers is negligible at cryogenic temperatures~\cite{Pobell2007, macdonald2015optical}. Optical frequency readout can furthermore relax space constraints in the cryostat by eliminating or reducing the need for cryogenic amplification and add channel capacity through the use of dense optical multiplexing~\cite{corcoran2020ultra}. Recent proof-of-principle experimental demonstrations of optical readout have used microwave-to-optics converters with three-dimensional optical~\cite{delaney2022superconducting} or microwave~\cite{arnold2023alloptical} cavities. However, three-dimensional transducer architectures for thousands of qubits exceed the spatial capacity of dilution refrigerators. Scalable optical readout of a superconducting qubit with large bandwidth, high repetition rate and a modular fiber-based approach has yet to be demonstrated.

Here we demonstrate a fully integrated optical readout system that upconverts the readout tone from a transmon qubit, enabling faithful measurement of the qubit state. We verify the operation of our optical readout system by measuring Rabi-oscillations between the qubit ground and excited state, as well as by performing Ramsey interferometry. Our device is completely independent from the qubit itself and has a modular coaxial connection to the qubit chip. It consists of a microwave-to-optics transducer formed by an optomechanical nanobeam cavity coupled to a piezoelectric transducer, described in ref.~\cite{weaver2022integrated}. The fully integrated device, requiring a footprint of less than 0.15 mm$^2$, allows for direct scaling of the transducer from an individual proof-of-concept device to multiple channels, in principle being able to read out over a thousand qubits in parallel with only a few optical fibers. 

\begin{figure}[h!]
	\centering
	\includegraphics[width = \linewidth]{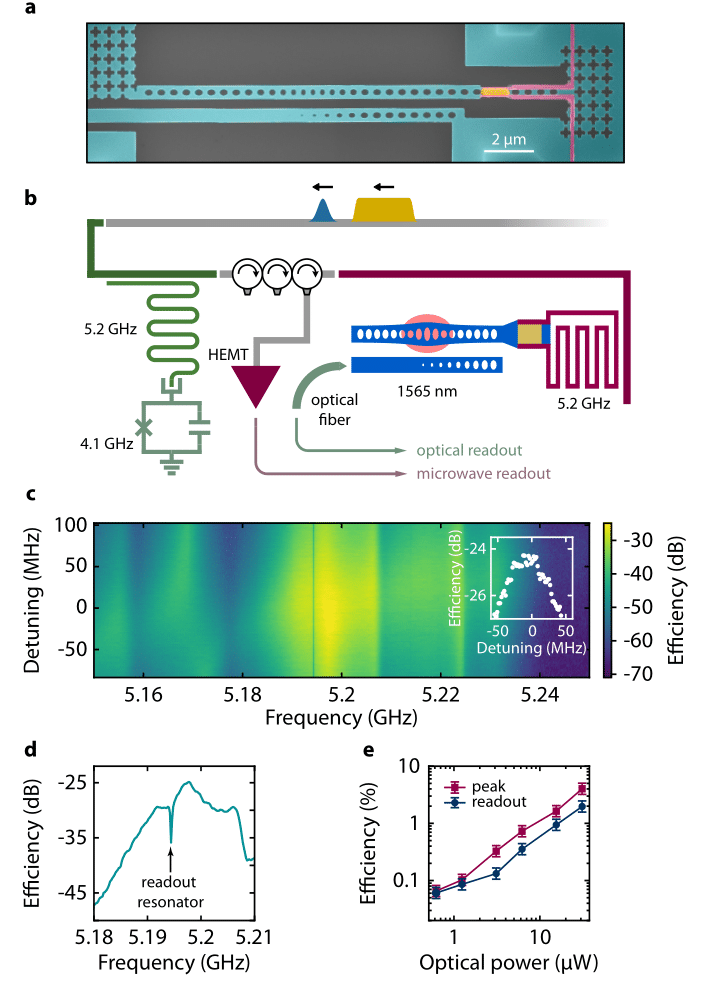}
	\caption{\textbf{Transducer performance.} 
		(a) Scanning electron microscope image of the piezo-optomechanical transducer comprising a piezoelectric block embedded in a superconducting resonator and an optomechanical photonic crystal cavity. (b) Illustration of the experimental configuration for qubit readout using a piezo-optomechanical microwave-to-optics transducer. Microwave readout (yellow) and control (blue) pulses used to operate a fixed frequency qubit are sent via coaxial cable into the transmission line of the transducer where the resulting qubit readout signal is upconverted to the optical domain. (c) Continuous-tone bidirectional transduction efficiency under application of $\SI{3.1}{\micro W}$ optical pump power as a function of signal frequency and of microwave resonator detuning from the peak transduction frequency, with the inset showing a vertical linecut crossing the point of optimum efficiency. A horizontal linecut is shown in (d), which exhibits a dip at $\SI{5.1944}{GHz}$ corresponding to the readout resonator of the qubit. The efficiency as a function of optical pump power is shown in (e), with the purple and blue lines representing the peak conversion efficiency and the efficiency at the qubit readout resonator frequency, respectively. The error bars represent the systematic errors.}
	\label{fig1}
\end{figure}

We use a fixed-frequency transmon qubit hosted on a quantum integrated circuit (QuIC) test device~\cite{RigettiFoundry} (qubit frequency $\omega_q/2\pi = \SI{4.07}{GHz}$), which is operated using microwave readout and control pulses, and dispersively coupled to a co-planar waveguide resonator (readout resonator frequency $\omega_r/2\pi = \SI{5.1944}{GHz}$). The output port of the single transmission line used to drive and read out the qubit is coupled to the microwave transmission line of a transducer chip via coaxial cables. A scanning-electron microscope image of a representative transducer device is shown in Fig.~\ref{fig1}a The transducer and qubit chips are physically separated by around $\SI{40}{cm}$ and the latter is magnetically shielded through multiple cryoperm casings, which also reduce the probability of stray light reaching the superconducting qubit. An illustration of the experimental configuration is shown in Fig.~\ref{fig1}b. Microwave readout signals are upconverted by a piezo-optomechanical transducer operated with an optical pump tone red-detuned from the optical resonance ($\lambda_o = \SI{1564.891}{nm}$, $\omega_o/2\pi = \SI{191.57}{THz}$) by the mechanical resonance frequency $\omega_m/2\pi = \SI{5.1944}{GHz}$. The optical pump tone is transmitted into the cryostat, where the light is sent towards the transducer via an edge coupler ($\SI{40}{\%}$ coupling efficiency). The total detection efficiency of the reflected optical signal is $\SI{17}{\%}$ (see the Supplementary Information section I for further details). As discussed in previous work~\cite{weaver2022integrated}, the transducer incorporates a field-tunable microwave resonator to (i) resonantly enhance the electromechanical interaction with a \ce{LiNbO3} piezoelectric block and (ii) facilitate efficient coupling to a $\SI{50}{\Omega}$ transmission line. 

Fig.~\ref{fig1}c shows the bidirectional microwave-to-optics conversion spectrum for different detunings of the transducer microwave resonator, under application of a continuous $\SI{3.1}{\micro W}$ red-detuned optical pump tone. The conversion efficiency is extracted using a four-port vector network analyzer as described in ref.~\cite{andrews2014bidirectional}. Electrical signals incident on the transducer microwave port are upconverted to telecom light via first the piezoelectric and subsequently the optomechanical interaction. This process modulates the microwave signal onto the optical pump, which is reflected from the optical port of the transducer. The resulting optical signal is demodulated outside of the cryostat using heterodyne detection. From the conversion spectrum, we extract a peak efficiency of $\SI{-24.8}{dB}$ ($\SI{0.33}{\%}$) at $\SI{5.1978}{GHz}$ with $\SI{3.1}{\micro W}$ of optical power, which is relatively insensitive to the detuning of the microwave resonator over a range of tens of MHz, as shown in the inset of Fig.~\ref{fig1}c. Fig.~\ref{fig1}d shows the efficiency as a function of the microwave drive frequency, exhibiting a $-\SI{3}{dB}$ transduction bandwidth of $\SI{4.7}{MHz}$, which is sufficiently large to facilitate state-of-the art qubit readout at MHz repetition rates~\cite{arute2019quantum, google2023suppressing}. The dip in the spectrum corresponds to the qubit readout resonator frequency, which is shifted by $\SI{3.5}{MHz}$ from the peak, leading to a slightly reduced efficiency of $\SI{-29.3}{dB}$ ($\SI{0.12}{\%}$). Due to the parametric nature of the optomechanical coupling rate~\cite{aspelmeyer2014cavity}, the transduction efficiency can be boosted by increasing the optical cavity photon number~\cite{mirhosseini2020superconducting} i.e., increasing the optical pump power, which is shown in Fig.~\ref{fig1}e. We observe a linear increase of efficiency for the pump powers considered, consistent with the expected small optomechanical cooperativity at these powers, and the large electromechanical cooperativity enabled by the piezoelectric interaction~\cite{xu2019frequency}. At the largest optical power of $\SI{31}{\micro W}$, we extract efficiency values of $\SI{4}{\%}$ and $\SI{2}{\%}$ at the peak and qubit readout resonator frequencies, respectively. For the latter, this gives a value of $\sim \SI{3e-3}{}$ for the overall quantum efficiency of the detection chain. 

\begin{figure}[h!]
	\centering
	\includegraphics[width = \linewidth]{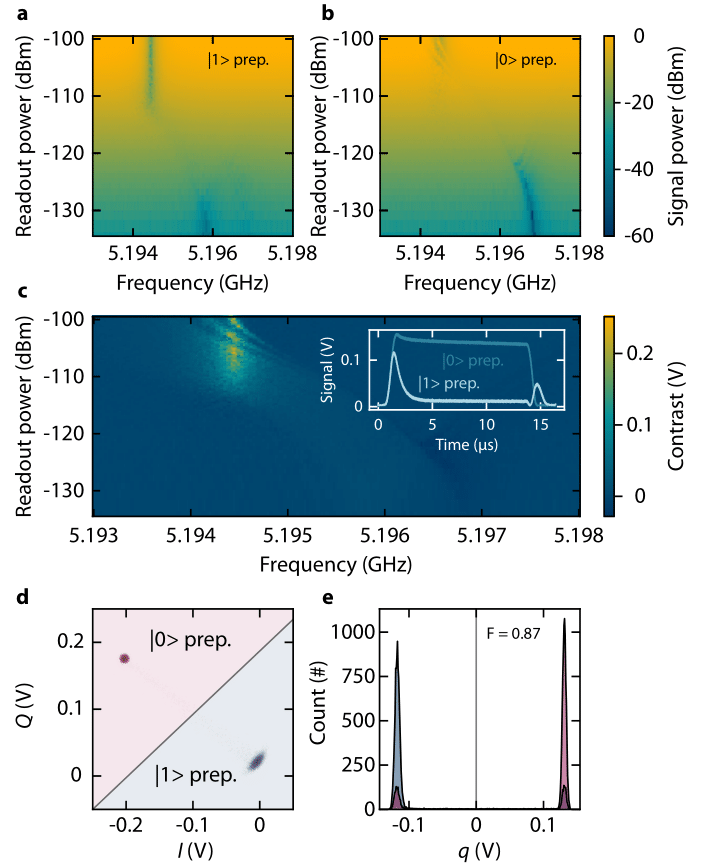}
	\caption{\textbf{Microwave qubit readout.} Readout signal magnitude for different readout frequencies and readout powers at the resonator with (a) the $\ket{1}$ and (b) the $\ket{0}$ state prepared. (c) Signal voltage difference obtained by subtracting (b) from (a). The inset shows averaged oscilloscope traces of the measured readout pulses, which have a duration of $\SI{14}{\micro s}$ and an integration window of $\SI{13.2}{\micro s}$ indicated by the vertical dashed lines to optimize the measured signal voltage difference between the two states. The largest difference is achieved at a readout power of $\SI{-105.8}{dBm}$ at the bare resonator frequency $\SI{5.1944}{GHz}$. Panels (d-e) show the statistical characterization of 10,000 single-shot readout measurements in (d) phase space and (e) binned by the distance $q$ between each point to a linear decision boundary.}
	\label{fig2}
\end{figure}

To assess the potential of this device for optical qubit readout, we consider the contribution of optical heterodyne shot noise, which adds half a quantum of noise at the end of the optical detection chain. In our current configuration, referred back to the transducer microwave port, this yields an input-referred added noise of approximately $\SI{2e3}{photons}$. This noise level impedes high-fidelity quantum non-demolition readout. However, it has been demonstrated that demolition readout can be used with high fidelity in transmon qubits~\cite{reed2010high, boissonneault2010improved, bishop2010response}. This approach exploits the occupation of higher level transmon states at large readout powers, which causes a qubit-state-dependence of the power at which the readout resonator enters the bare regime~\cite{boissonneault2010improved}. At such high readout powers, the signal is large enough to overcome typical amplifier noise at 4 K, enabling single-shot qubit readout without the use of a quantum-limited amplifier~\cite{reed2010high}. 

To define a benchmark for optical readout with a transducer, we proceed by characterizing the high-power readout fidelity with microwave signals only, using a $\SI{4}{K}$ high-electron-mobility-transistor (HEMT) amplifier as the first amplifier in the microwave signal-processing chain. The microwave input readout signal is generated by upconverting signals from a commercially available base-band Qblox Qubit Readout Module (QRM) to the RF band with an I-Q mixer and delivered to the qubit transmission line~\cite{Qblox2024}. The returned readout signal is interpreted through subsequent downconversion using an I-Q mixer, digitization and digital demodulation. Figs.~\ref{fig2}a-b show the recorded microwave signal magnitude as a function of the frequency and amplitude of $\SI{14}{\micro s}$-long readout pulses with the $\ket{1}$ and $\ket{0}$ qubit states prepared, respectively. Optimal signal voltage difference is achieved at the bare readout resonator frequency at a cavity drive power of $\SI{-105.8}{dBm}$ (see Fig.~\ref{fig2}c). This readout power corresponds to a field amplitude in the readout resonator of $ \sqrt{n_{cav}} \simeq \SI{90}{photons^{1/2}}$. The inset of Fig.~\ref{fig2}c shows averaged oscilloscope traces of the detected readout signals at the point of optimal signal voltage difference. To obtain maximum difference between the two states, we choose a $\SI{13.2}{\micro s}$ integration window, as indicated by the dashed vertical lines. Fig.~\ref{fig2}d shows the statistics of 10,000 single-shot readout measurements in phase space. We optimize a linear decision boundary to determine the experimental readout fidelity defined as $F = 1-\qty[p(1|0)+p(0|1)]/2$, which includes both state preparation and readout fidelity and serves as a lower bound for the readout fidelity (see Fig.~\ref{fig2}e). We extract a fidelity of $F = \SI{87}{\%}$, corresponding to an error rate of $\SI{6.5}{\%}$, which we attribute to uncontrolled switching events. Such events can occur due to suboptimal thermalization of the qubit chip, thermally out-of-equilibrium quasiparticles ~\cite{wenner2013excitation, geerlings2013demonstrating, jin2015thermal}, or due to qubit energy relaxation during the readout pulse. From the number of photons in the readout pulse $\sim \SI{1.1e5}{}$ and the signal-to-noise ratio $\text{SNR} \approx \SI{78}{}$, given by the ratio of the separation and the standard deviation of the statistical distributions, we identify a microwave added noise level of $\sim \SI{17}{photons}$, which can largely be attributed to HEMT noise ($\sim\SI{7}{photons}$) in combination with $\SI{3}{dB}$ of loss introduced by the side-coupled qubit readout resonator.

\begin{figure}[h!]
	\centering
	\includegraphics[width = \linewidth]{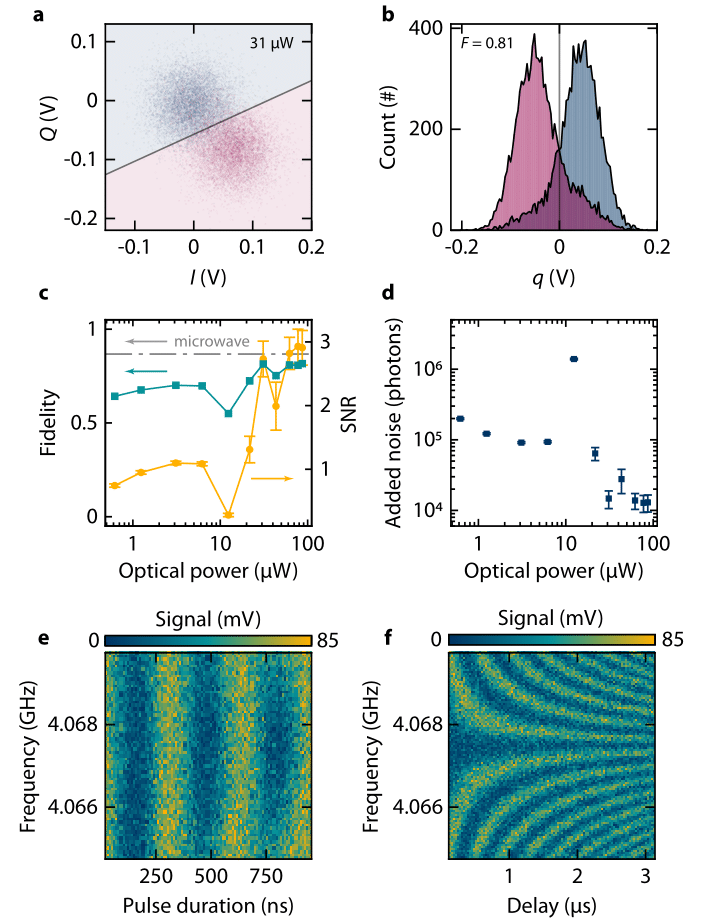}
	\caption{\textbf{Optical qubit readout.} (a) Statistical characterization of 10,000 single-shot optical readout measurements using $\SI{31}{\micro W}$ of optical pump power. (b) The data from (a) binned by the distance to a linear decision boundary $q$. (c) Optical readout fidelity (left axis) and signal-to-noise ratio (right axis) as a function of optical pump power. The dashed line highlights the microwave readout fidelity. The dip at $\SI{12.4}{\micro W}$ is attributed to a transition in the optically-dependent phase response of the microwave resonator of the transducer (see Supplementary Information section II). Panel (d) shows the input-referred added noise as a function of optical pump power in units of photons. All error bars are s.d. Panels (e-f) show a Rabi chevron pattern and Ramsey fringes, respectively, measured using optical readout.}
	\label{fig3}
\end{figure}

Having optimized the high-power readout pulse and having set a benchmark for its performance, we now perform high-power optical readout using the configuration depicted in Fig.~\ref{fig1}b. The $\SI{31}{\micro W}$ optical pump is pulsed in unison with the readout pulse using an acousto-optic modulator (AOM). Between shots, a pause of $\SI{250}{\micro s}$ is inserted to allow for the qubit to reset through longitudinal relaxation ($T_1 \approx \SI{60}{\micro s}$), yielding an average power delivered into the cryostat of $\SI{1.7}{\micro W}$ at a $\SI{5.3}{\%}$ duty cycle. Using the same procedure as for microwave-only readout, we recover a single-shot optical readout fidelity of around $\SI{81}{\%}$ after 10,000 measurements (see Figs.~\ref{fig3}a-b). Fig.~\ref{fig3}c shows the readout fidelity and signal-to-noise ratio, determined from bimodal Gaussian fits, as a function of optical power. We find that at low to moderate optical powers, the experimentally determined fidelity is close to the theoretically maximum achievable value of $\SI{70}{\%}$ for a SNR near unity~\cite{chen2023transmon}. At high optical power, the experimental fidelity of $\SI{82}{\%}$ falls short of the theoretical maximum of $\sim \SI{93}{\%}$ for an SNR of $\sim 3$, suggesting a significant contribution of uncontrolled switching events. However, the SNR of the optical readout is lower than that of the microwave-only readout by a factor of $\sim 25$ due to a considerably larger added noise value. Fig.~\ref{fig3}d shows the added noise as a function of optical power, which decreases from $\sim \SI{2e5}{photons}$ to $\sim \SI{e4}{photons}$ as the optical power is increased from $\SI{0.6}{\micro W}$ to $\SI{87}{\micro W}$. We estimate the heterodyne shot noise contribution to be around $\SI{2e3}{photons}$, which dominates over thermal noise. However, an additional noise source, possibly shot noise introduced by the optical pump, ultimately limits the single-shot readout fidelity. We refer the reader to the Supplementary Information section III for a detailed breakdown and analysis of the added noise. Nevertheless, we find that our optical readout technique is quite suitable for canonical qubit characterizations such as Rabi spectroscopy and Ramsey interferometry (see Figs.~\ref{fig3}e-f).

\begin{figure}[h!]
	\centering
	\includegraphics[width = \linewidth]{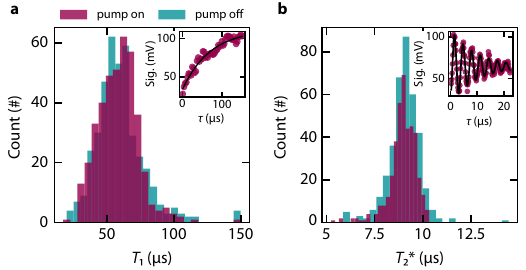}
	\caption{\textbf{Absence of pump induced decoherence.} (a) Statistical distributions of 401 $T_1$ measurements, comparing the decay times with a $\SI{3.1}{\micro W}$ optical pump on (purple) and off (cyan) yielding $\SI{60.2 \pm 3.6}{\micro s}$ versus $\SI{58.9 \pm 3.0}{\micro s}$. Similarly, panel (b) compares the statistical distributions of 401 $T_2^*$ measurements with the pump on (purple) and off (cyan) yielding $\SI{8.97 \pm 0.17}{\micro s}$, versus $\SI{9.03 \pm 0.14}{\micro s}$. The insets in both panels show respective example curves with the optical pump on.}
	\label{fig4}
\end{figure}

At large optical pump powers, thermal microwave photons emitted from the transducer may cause qubit decoherence. In a recent work, it has been shown that transverse qubit decay can be mitigated through implementation of sufficient isolation from a microwave transducer pump~\cite{delaney2022superconducting}. We therefore proceed to characterize qubit longitudinal and transverse decay times by measuring $T_1$ and $T_2^*$ with $\SI{60}{dB}$ of isolation between the transducer and qubit readout resonator. The transducer microwave reflection output is sent to a HEMT microwave amplifier at 4K and used to determine the $T_1$ and $T_2^*$ decay times using microwave-only readout, comparing scenarios with an optical transducer pump of $\SI{3.1}{\mu W}$ continuously on or off. To obtain a robust estimation, we perform 401 measurements for each scenario. Fig.~\ref{fig4}a shows the distribution of the extracted $T_1$ times, with an inset illustrating an example curves and corresponding fit. We find no statistical difference in the mean $T_1$ with the pump off ($\SI{60.2 \pm 3.6}{\micro s}$) and the pump on ($\SI{58.9 \pm 3.0}{\micro s}$). Likewise, in Fig.~\ref{fig4}b no statistical difference in the mean $T_2^*$ with pump off ($\SI{9.03 \pm 0.14}{\micro s}$) and pump on ($\SI{8.97 \pm 0.17}{\micro s}$) can be observed. With reduced isolation between the qubit and the transducer, emitted microwaves from the transducer do have an impact on the qubit coherence and the variance of $T_1$ and $T_2^*$ (see Supplementary Information section IV), which has also been a major limitation for quantum transduction when placing the qubit on the same chip as the transducer~\cite{mirhosseini2020superconducting}. However, $\SI{60}{dB}$ of isolation is sufficient to prevent backaction of the transducer on the qubit, highlighting the importance of a modular approach of qubit and transduction technology for practical scaling of quantum computers.

Optical readout through a transducer and a fiber opens up new possibilities for configuration of the readout chain in fridges. In particular, a HEMT amplifier dissipates typically close to $\SI{10}{mW}$ of power, which constitutes a significant thermal load on the 4K plate. The microwave-to-optics transducer dissipates a maximum power of $\SI{31}{\micro W}$, which is further decreased by the readout duty cycle. The transducer could also be placed on the still of the dilution refrigerator, dissipating only a small fraction of the available cooling power~\cite{krinner2019wiring} and completely removing the cabling and amplifier heat loads from higher stages. One possible extension is to use transducers as a source for microwave control signals~\cite{warner2023coherent} or for the generation as well as detection of readout signals, as demonstrated in recent work~\cite{arnold2023alloptical}. Decreasing the isolation to admit an excitation signal may decrease the $T_2^*$ of the qubit while the transducer optical pump is on, but the pulsed approach used here would protect the qubit during operations. A space efficient solution could be to include on-chip Purcell filters~\cite{reed2010fast, jeffrey2014fast, sete2015quantum} and isolators~\cite{chapman2017widely} and to use small form factor coaxial flex-cabling between stages and modules~\cite{smith2024cabling, monarkha2024cabling}. 

Added noise in our current setup prevents us from fully matching the microwave-only readout performance. Extra local oscillator photons causing this noise can, however, be removed through optical filtering outside the dilution refrigerator, reducing the added noise by $\sim 7.7$ dB and enabling the full potential of our optical qubit readout scheme. Additionally, we anticipate that better frequency matching between the transduction frequency and the qubit readout resonator, as well as improvement in transducer cooperativity matching will improve the transduction efficiency by 3 dB and 6 dB, respectively and accordingly the added noise by 10 dB. Finally, adding in a quantum limited amplifier, such as a traveling wave parametric amplifier or Josephson parametric amplifier could further reduce the noise by at least 20 dB~\cite{krinner2019wiring}. In the future, alternative transducer geometries such as optically-overcoupled two-dimensional photonic crystals could reduce the optical pump power and the thermal noise such that isolation from the qubit is no longer required~\cite{sonar20242dcavities,chen20242dcavities}. These improvements will result in an optical readout system with close to quantum limited performance which could perform non-demolition qubit readout with high fidelity and shorter pulse lengths, while at the same time significantly reducing the heat load on the dilution refrigerator compared to all-microwave readout approaches.

In summary, we have demonstrated high-power optical qubit readout using a scalable integrated piezo-optomechanical transducer with a single-shot fidelity of $\SI{81}{\%}$. These transducers could be arrayed together to enable readout of thousands of superconducting qubits in a cryogenic environment, expanding the size of quantum processors in a single dilution refrigerator.

\medskip

\section{Acknowledgments}

We gratefully acknowledge assistance from Emma He and the hospitality of the Department of Quantum Nanoscience at Delft University of Technology, as well as the Kavli Nanolab Delft. QphoX would also like to thank the European Innovation Council (EIC Accelerator QModem 190109269) for financial support. Qblox acknowledges support from the European Commission under Grant agreement 969201.

\section{Author contributions}

T.C.T., M.J.W., M.Z., F.H., R.S. and S.G. designed the experiment. T.C.T., M.J.W., P.D., M.Z., A.C.B., C.F. and E.C. collected and analysed the data. F.B., P.D., M.L., K.L.S and R.S. designed and fabricated the transducer device. F.K.V., C.C.B. and J.C.O. designed electronic hardware for qubit operation and data acquisition. E.L., M.F., Y.M. and J.Y.M. provided qubit and mounting hardware, as well as experimental support. T.C.T., M.J.W., R.S. and S.G. wrote the manuscript with input from all authors.

\section{Competing interests}

T.C.T., M.J.W., F.B., P.D., M.L., K.L.S., M.Z., F.H., A.C.B., C.F., E.C., R.S and S.G. are or have been employed by QphoX B.V. and are, have been, or may in the future be participants in incentive stock plans at QphoX B.V. F.K.V., C.C.B. and J.C.O. declare a financial interest in Qblox B.V. J.Y.M., M.F., E.L., Y.M. are or have been employed by Rigetti \& Co, LLC.  J.Y.M., M.F., E.L., Y.M. are, have been, or may in the future be participants in incentive stock plans at Rigetti \& Co, LLC.

\section{Data availability}
Source data for the figures are available via Zenodo at \href{https://doi.org/10.5281/zenodo.14293252}{https://doi.org/10.5281/zenodo.14293252}.

\setcounter{figure}{0}
\renewcommand{\thefigure}{S\arabic{figure}}
\setcounter{equation}{0}
\renewcommand{\theequation}{S\arabic{equation}}

\clearpage

\section*{Supplementary Information}
\label{SI}

\section{Fabrication}

\begin{figure}[ht!]
	\centering
	\includegraphics[width = 1\linewidth]{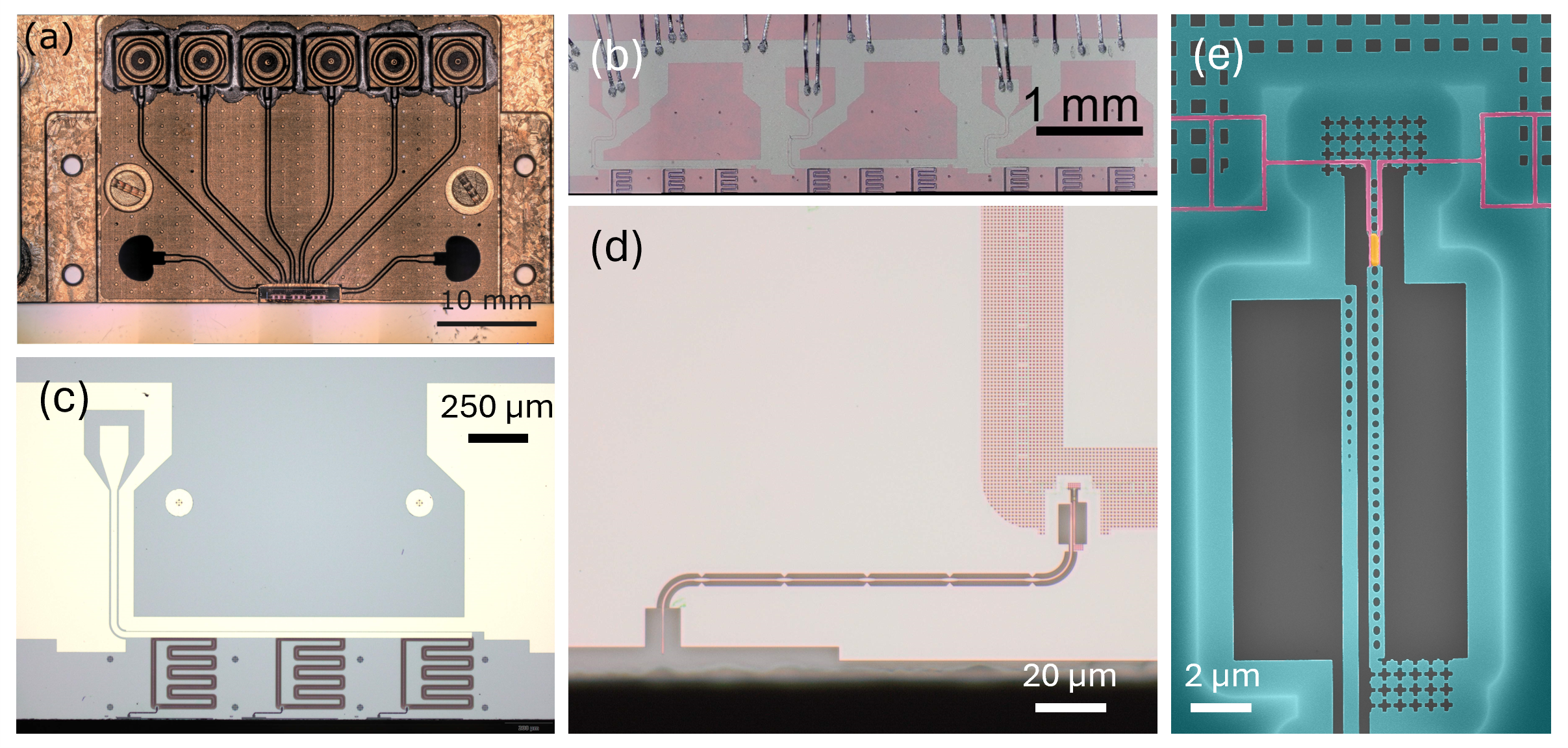}
	\caption{\textbf{Integrated piezo-optomechanical transducer.} Optical microscope images of (a) the chip carrier with a mounted transducer chip, (b) the transducer chip with wirebonds to the chip carrier SMP electrical ports, (c) three electrically multiplexed transducer devices coupled to a microwave transmission line in a hanger geometry, (d) the transducer optical waveguide edge coupler and (e) a SEM image of a representative transducer device.}
	\label{fig:pictures}
\end{figure}

The starting material for these devices is a \SI{330}{nm} film of X-cut \ce{LiNbO3} (LN) bonded on an high resistivity silicon on insulator (SOI) substrate. The full device fabrication consists of four main parts. In the first one, an \ce{Ar} milling process is patterning the LN layer to form small blocks. The second one involves the formation of the nanobeams using a reactive-ion etch (RIE) step. Afterwards, a deposition of a \SI{45}{nm} thick \ce{MoRe} layer fabricates the bondpads, the feedline, and the resonators. Finally, a buffered oxide etch (BOE) solution selectively etches the sacrificial \ce{SiO2} layer underneath the Si device layer to suspend the devices. This step enables the oxide under the resonators to be removed and thus to increase the electrical quality factor of the devices. Further details regarding the fabrication is described in previous work~\cite{weaver2022integrated}.

After fabrication, samples are mounted on top of a copper chip carrier using GE varnish and the transducer microwave ports are wire bonded to electrical lines ending in soldered SMP connectors (Figs.~\ref{fig:pictures}a-b). Each microwave transmission line electrically multiplexes three transducer devices in hanger geometry in reflection (Fig.~\ref{fig:pictures}c). The optical port of the transducer device consists of a waveguide edge coupler at the edge of the chip, which can be coupled to with a lensed fiber (Fig.~\ref{fig:pictures}d). A SEM image of a representative transducer device is shown in (Fig.~\ref{fig:pictures}e).  

\section{Transducer characterization}

\begin{table*}[ht]
	\centering
	\caption{Device and experimental parameters.}
	\begin{tabular}{c|l|c|c|c}
		Parameter Name & Description & Value & Error & Units \\ \hline
		Transducer Property & & & & \\ \hline
		$\omega_m/2\pi$ & mechanical resonance frequency & 5.19442 & & GHz \\
		$\omega_o/2\pi$ & optical resonance frequency & 191.57 & & THz \\
		$\omega_p/2\pi$ & frequency of peak conversion efficiency & 5.198 & & GHz \\
		$\kappa_m/2\pi$ & mechanical linewidth & 1.53 & 0.01 & MHz \\
		$\kappa_e/2\pi$ & total microwave loss rate & 23.6 & 5.32 & MHz \\
		$\kappa_{ee}/2\pi$ & microwave external coupling rate & 12.2 & 2.84 & MHz \\
		$\kappa_{ei}/2\pi$ & microwave internal loss rate & 11.4 & 2.48 & MHz \\
		$\kappa_{o}/2\pi$ & total optical loss rate & 5.16 & 0.03 & GHz \\
		$\eta_e \equiv \frac{\kappa_{ee}}{\kappa_e}$ & microwave resonator coupling & 0.517 & 0.08\\
		$\eta_o \equiv \frac{\kappa_{oe}}{\kappa_o}$ & optical resonator coupling & 0.5 & 0.18 &\\ \hline 
		Qubit Property & & & & \\ \hline
		$\omega_r/2\pi$ & bare readout resonator frequency & 5.194 & & GHz\\
		$\omega_{r}'/2\pi$ & dressed readout resonator frequency & 5.198 & & GHz\\
		$\omega_q/2\pi$ & qubit frequency & 4.07 & & GHz\\
		$\kappa_{re}/2\pi$ & readout resonator linewidth & 500 & & kHz \\
		$\kappa_{ree}/2\pi$ & readout resonator external coupling rate & 450 & & kHz \\
		$\kappa_{ree}/2\pi$ & readout resonator internal coupling rate & 50 & & kHz \\
		$g/2\pi$ & qubit-resonator coupling & 52 & 0.1 & MHz\\
		$\chi/2\pi$ & dispersive shift & 512 & 20 & kHz\\
		\hline Experimental Setup Property & & & & \\ \hline
		$\eta_{fiber}$ & fiber coupling efficiency & 0.40 & 0.01 &\\
		$\eta_{od}$ & optical output detection efficiency & 0.43 & 0.05 & \\
		$\eta_{tod}$ & total optical detection efficiency & 0.17 & 0.05 & \\
	\end{tabular} 
	\label{tab:table}
\end{table*}

\begin{figure}
	\centering
	\includegraphics[width = 1\linewidth]{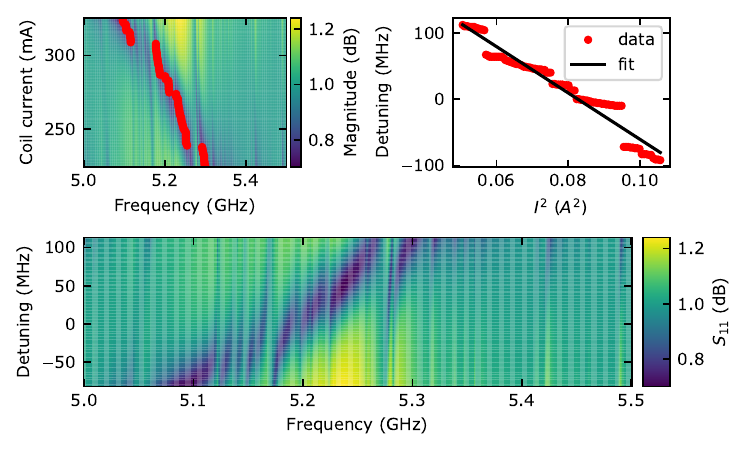}
	\caption{\textbf{Tuning of the microwave resonance.} (a) Microwave reflection spectrum as for different coil currents and probe frequencies. The red dots indicate the microwave resonance dip. (b) Microwave resonance as a function of the square of the coil current with a linear fit. (c) Microwave reflection spectrum as a function of probe frequency and resonance detuning.}
	\label{fig:coil_tuning}
\end{figure}

The (side-coupled) transducer microwave resonator consists of a $\SI{160}{nm}$ wide superconducting wire, organized into a half-wave laddered comb-like loop. The ladder subloops combined with the high kinetic inductance (around $\SI{10}{pH/square}$) of the \ce{MoRe} superconducting layer allow for tuning of the resonance frequency, as described in ref.~\cite{xu2019frequency}. An external superconducting coil is mounted in close proximity to the transducer chip, yielding a field-to-current ratio of about $\SI{10}{mTA^{-1}}$ at the sample. This leads to a tunability of typically several hundreds of $\si{MHz}$ and facilitates bringing the microwave resonator into resonance with the transducer optomechanically-active electromechanical mechanical mode. Fig.~\ref{fig:coil_tuning} illustrates tuning of the microwave resonance through several electromechanical modes.

\begin{figure}
	\centering
	\includegraphics[width = 1\linewidth]{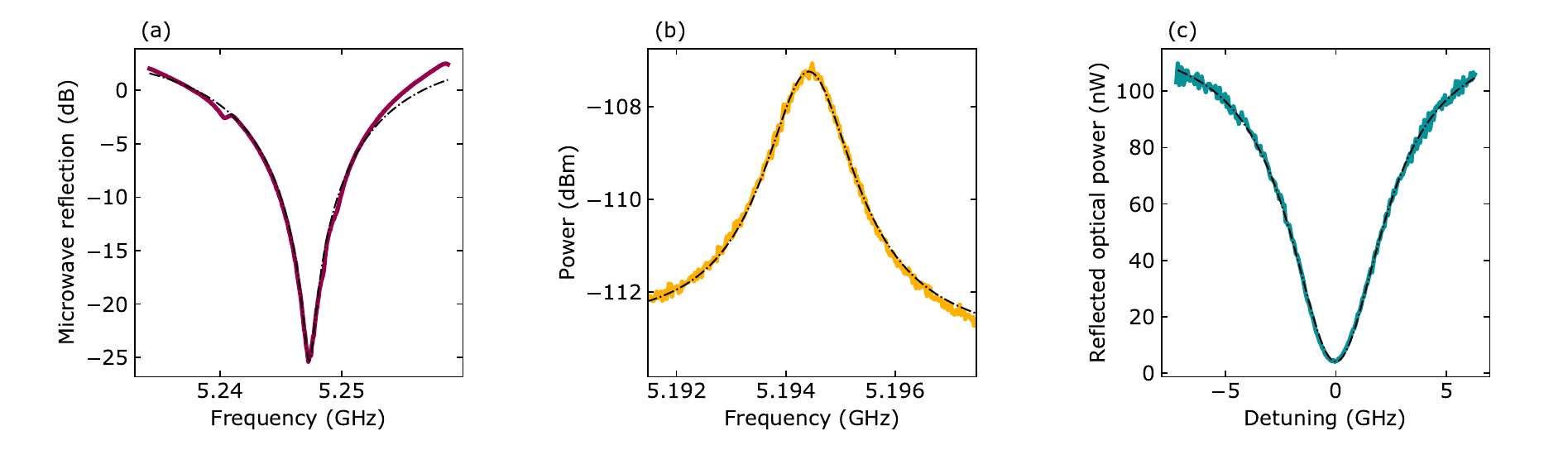}
	\caption{\textbf{Resonances of the piezo-optomechanical transducer}. (a) Microwave resonance with the magnetic field set to a detuning far away from any electromechanically active mechanical modes. (b) Thermal modulation of the optical frequency due to the optomechanically active mechanical mode, measured through optical heterodyne detection. (c) Reflection measurement of the optical resonance of the photonic crystal cavity.}
	\label{fig:resonances}
\end{figure}

Fig.~\ref{fig:resonances} shows the electrical, mechanical and optical resonances of the piezo-optomechanical system. To determine the external and internal electrical coupling rates, the microwave resonator is detuned from any electromechanical modes and its transmission spectrum is fit following the routine described in ref.~\cite{rodrigues2019coupling} (see Fig.~\ref{fig:resonances}a). We find internal and external coupling rates of $\SI{11.4}{MHz}$ and $\SI{12.2}{MHz}$, respectively i.e., an overcoupled resonator with a (side-)coupling coefficient of $\eta_e = 0.517$. The mechanical resonance of the optomechanical cavity is determined at $\SI{4}{K}$, where it exhibits ample thermal motion. The GHz modulation of the hundreds of THz optical resonance by the thermally-excited mechanical mode is extracted through optical heterodyne detection, using a high-speed photodetector. The result is shown in Fig.~\ref{fig:resonances}b. We fit the data with a Lorentzian function, which yields a resonance frequency of $\omega_m/2\pi = \SI{5.19442}{GHz}$ with a total linewidth of $\kappa_m/2\pi = \SI{1.53}{MHz}$. Note that, due to hybridization with the electromechanical mode of the \ce{LiNbO3} piezoelectric block, the frequency of peak transduction efficiency ($\omega_p/2\pi = \SI{5.198}{GHz}$) is slightly shifted from the mechanical resonance. Fig.~\ref{fig:resonances}c shows the optical cavity response, exhibiting a frequency of $\omega_o/2\pi=\SI{191.57}{THz}$, linewidth of $\kappa_o/2\pi = \SI{5.16}{GHz}$ and waveguide-to-resonator coupling efficiency of $\eta_o \approx 0.5$, with a fiber-to-waveguide coupling of $\eta_{fiber} = 0.4$. The results are summarized in Table~\ref{tab:table}.

Following the approach of ref.~\cite{andrews2014bidirectional}, we characterize the transducer conversion efficiency with a four-port vector network analyzer (VNA) measurement. Upconversion is measured by applying a continuous microwave tone to the transducer electrical input port and demodulating the transduced optical signal using a heterodyne scheme incorporating a fast photodetector. The resulting microwave signal is collected by one of the two VNA input ports. Downconversion is measured through application of a microwave tone to an electro-optic modulator (EOM), generating optical sidebands on the red-detuned optical pump directed towards the optical input of the transducer. The transducer demodulates the composite optical signal to the microwave domain. The signal is subsequently amplified by $\SI{4}{K}$ HEMT and room-temperature low-noise amplifiers and collected by the other VNA input port. The transduction efficiency is calculated as

\begin{align}
	\eta = 2\alpha \frac{\abs{S_{eo}} \abs{S_{oe}}}{\abs{S_{ee}}\abs{S_{oo}}},
\end{align}

\noindent where $S_{eo}$ ($S_{oe}$) is the upconversion (downconversion) and $S_{ee}$ ($S_{oo}$)is the reflected microwave (optical) signal. The term $2\alpha \approx 1.3$ corrects for a nonzero rejection of the lower optical sideband. The data reported here was recorded with a phase EOM, with the correction to the efficiency values calibrated by a subsequent amplitude EOM transduction measurement.

\begin{figure}
	\centering
	\includegraphics[width = 1\linewidth]{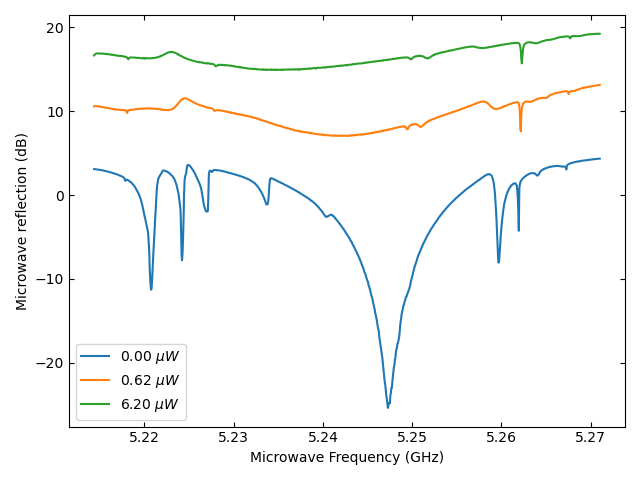}
	\caption{\textbf{Optical effect on the transducer microwave resonator}. Microwave resonance of the tranducer measured in reflection with (blue) no optical pump power, (orange) $\SI{0.62}{\mu W}$ of optical pump power and (green) $\SI{6.2}{\mu W}$ of optical pump power. Curves are offset vertically for clarity}
	\label{fig:opt_microwaveres}
\end{figure}

In the same way as we observed in reference~\cite{weaver2022integrated}, the optical field required to operate the transducer results in an increase loss rate of the superconducting microwave resonator, as well as a reduction of the resonator frequency. Figure \ref{fig:opt_microwaveres} displays the reflected microwave power around the microwave resonator under increasing optical power input into the system. The data are taken for an pump red-detuned from the optical resonance by 5.2 GHz, as used for the transducer characterization and qubit readout in the main text.

\begin{figure}
	\centering
	\includegraphics[width = 1\linewidth]{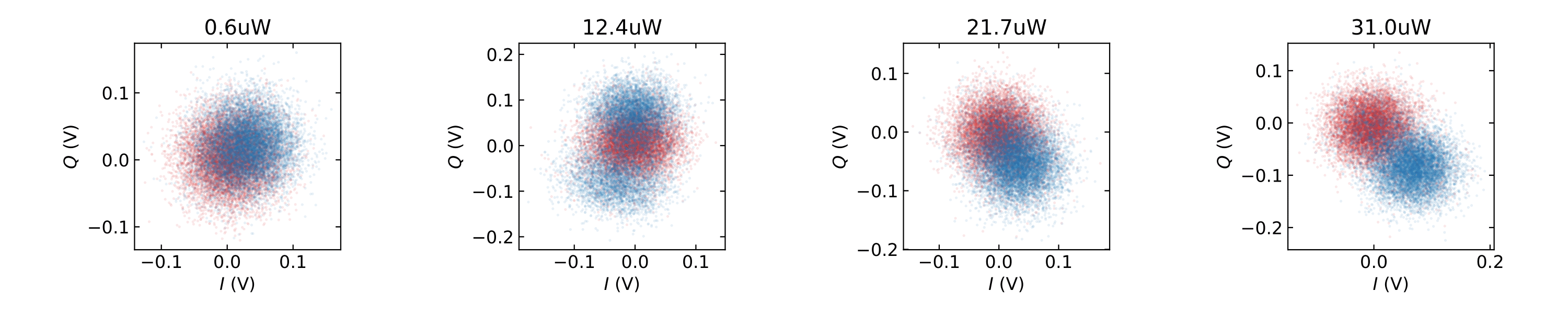}
	\caption{\textbf{Optically-dependent readout phase shift.} Single-shot optical readout statistical distributions for various optical powers with (red) the $\ket{1}$ state and (blue) the $\ket{0}$ state prepared.}
	\label{fig:fidelity_phase}
\end{figure}

Fig.~\ref{fig:fidelity_phase} illustrates the impact of the optical pump on the phase response of the optical readout system. The phase space statistical distributions of the single-shot readout experiments exhibit a phase shift with high optical powers. At the transition between low and high powers at 12.4 uW the system is unstable and jumps between two positions in phase space. This explains the low fidelity at this power in Fig. 3c of the main text. We attribute this phase shift to the previously observed dependence of the transducer microwave resonator amplitude and phase response on optical pump power~\cite{weaver2022integrated}.

\section{Added noise}

In the main text we report added noise, which is the noise referred to the beginning of the microwave field detection path, immediately after the qubit readout resonator. Laser amplitude fluctuations are mitigated by a balanced heterodyne scheme, which largely suppresses ($\sim \SI{40}{dB}$) any common mode noise between the signal and local-oscillator optical paths. The overall noise floor is dominated by Poissonian shot noise of the powerful local oscillator ($\sim \SI{1}{mW}$), which amplifies the coherent tones from the signal input port and adds shot noise, both proportional to the square root of the optical power. The input-referred added noise therefore, is mainly determined by the transduction efficiency and the optical detection efficiency in the signal path. The two main noise contributions in our experimental setup are thermal noise in the transducer mechanical resonator and optical heterodyne shot noise. In the following sections we briefly discuss the extraction of these two quantities. 

\subsection{Thermal noise}
\label{sec:added_thermal_noise}

The optical power resulting from the upconversion of a given microwave input power $P_{e, in}$ is given by

\begin{align}
	P_{o, sig} =  \frac{\omega_o}{\omega_e} \eta_t \eta_c P_{e, in}, 
\end{align}

with $\omega_e$ ($\omega_o$) the microwave (optical) angular frequency, $\eta_t$ the photon conversion efficiency and $\eta_c$ the optical collection efficiency, comprising optical losses along the chain, as well as the quantum efficiency of the photodetector. Thermal noise power added by the mechanical resonator $P_{m, th}$ appears on the output as

\begin{align}
	P_{o, th} = \frac{\omega_o}{\omega_m} \eta_{om} \eta_c P_{m, th}  = \frac{\omega_o}{\omega_m} \eta_t \eta_c P_{e, th},
\end{align}

where we use the relation $\eta_t = \eta_{om} \eta_{em}$, with $\eta_{om}$ ($\eta_{em}$) the optomechanical (electromechanical) conversion efficiency. The optical signal-to-thermal-noise ratio is then given by

\begin{align}
	\text{SNR}_o = \frac{\omega_m}{\omega_e} \frac{P_{e, in}}{P_{e, th}} \approx \frac{P_{e, in}}{P_{e, th}},
\end{align}

with $\omega_m$ the mechanical resonance frequency and $P_{e, th}$ the input-referred thermal noise power. In a heterodyne scheme, the measured signal-to-noise ratio on a photodiode output is linearly proportional to the optical signal-to-noise ratio on the input. We then have for the thermal noise equivalent power

\begin{align}
	P_{e, th} = \frac{P_{e, in}}{\text{SNR}_{meas}},
	\label{eq:thermal}
\end{align}

which is in general a frequency dependent property due to the resonant nature of the thermal noise. Accordingly, the added noise in an experiment depends sensitively on the microwave signal and mechanical resonance frequencies, with the thermal noise being smaller or even negligible when they differ and maximum when they are equal. 

\begin{figure}[h]
	\centering   
	\includegraphics[width = 1.\linewidth]{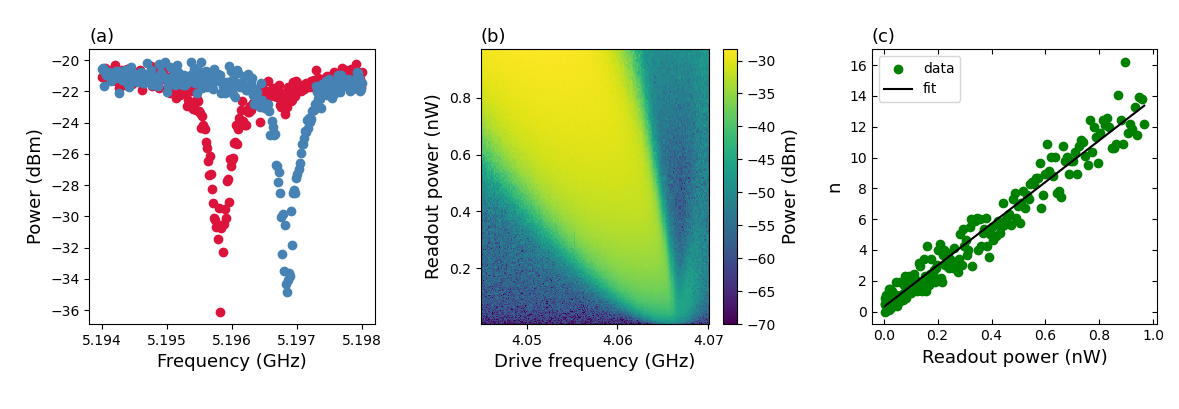}
	\caption{\textbf{Line attenuation}. Panel (a) shows the readout resonator response with the qubit prepared in the (blue) $\ket{0}$ state and (red) $\ket{1}$ state, from which we extract a dispersive shift $\chi/2\pi = \SI{512}{kHz}$. Panel (b) depicts two-tone spectroscopy data as a function of room temperature readout power and qubit drive frequency for fixed qubit drive power. The data shows the Stark shift of the qubit frequency due to an increase in the intracavity photon number of the readout resonator. The extracted photon number as a function of room temperature readout power is shown in panel (c).}
	\label{fig:line_attenuation}
\end{figure}

To determine the microwave power on the input of the transducer microwave port, we make use of the AC Stark shift effect, which leads to a readout power-dependent frequency shift of the qubit given by $\Delta \nu = 2\bar{n} \chi$, with $\chi$ the dispersive shift~\cite{schuster2005ac} and $\bar{n}$ the intracavity photon number of the readout resonator. Figure \ref{fig:line_attenuation}a shows the readout resonator response for the qubit prepared in the $\ket{0}$ and $\ket{1}$ states, revealing a dispersive shift of $\chi/2\pi = \SI{512}{kHz}$. If the readout resonator is driven on resonance, the intracavity photon number is given by~\cite{aspelmeyer2014cavity, kalaee2019quantum}

\begin{align}
	\bar{n} = \frac{4}{\hbar \omega_0} \frac{\kappa_{ext}}{\kappa_{tot}^2} P_{e, in} = \frac{4}{\hbar \omega_0} \frac{\kappa_{ext}}{\kappa_{tot}^2} \alpha P_{e, rt} ,
\end{align}

with $\kappa_{tot}$ ($\kappa_{ext}$) the total (external) loss rate of the readout resonator, $\omega_0$ the unshifted dressed readout resonator frequency, $\alpha$ the cryogenic line attenuation and $P_{e, rt}$ the microwave power applied at room temperature. The qubit frequency as a function of readout power is determined by two-tone spectroscopy (see Fig.~\ref{fig:line_attenuation}b). From the power-dependent frequency shift, we determine the intracavity photon number $\bar{n}$ as a function of the microwave readout power applied at room temperature (see Fig.~\ref{fig:line_attenuation}c). With the readout resonator loss rates $\kappa_{ext}/2\pi = \SI{0.45}{MHz}$ and $\kappa_{tot}/2\pi = \SI{0.5}{MHz}$, we extract an attenuation of $\SI{73.8}{dB}$. We estimate the total line attenuation to be $\SI{74.5}{dB} \pm \SI{0.5}{dB}$ due to additional insertion losses of circulators and connectors between the qubit readout line and the transducer microwave port.

\begin{figure}[h]
	\centering
	\includegraphics[width = 1\linewidth]{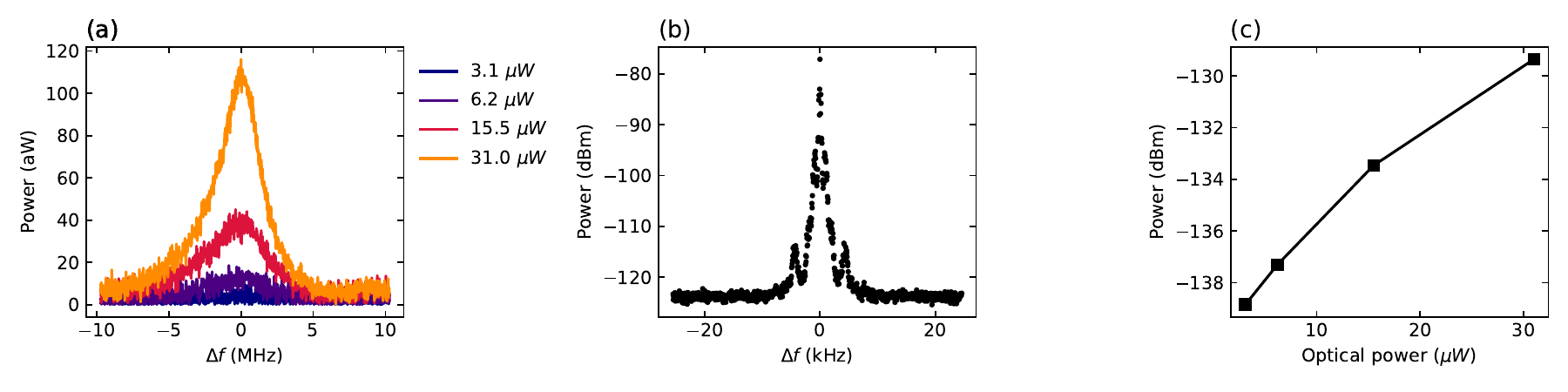}
	\caption{\textbf{Thermal noise.} (a) Background corrected resonant thermal noise spectra measured for different optical powers. (b) Upconverted microwave tone at $\SI{15.5}{\micro W}$ of optical power. (c) Thermal noise power as a function of optical power.}
	\label{mw_line_calibration}
\end{figure}

We determine the thermally added noise by measuring simultaneously the upconverted signal of a applied microwave tone and thermal phonons in the optomechanical resonator at $\SI{15.5}{\micro W}$ of applied pump power (see Fig.~\ref{mw_line_calibration}a) using heterodyne detection. Correcting for non-thermal background noise sources and assuming a Lorentzian line shape of the thermal noise, we extract an average power value of $\SI{-133.7}{dBm}$ over a bandwidth of $\SI{1.5}{MHz}$. To determine the thermal noise equivalent power, we extract the signal-to-noise ratio on the photodiode output between the upconverted microwave calibration tone and the thermal noise contribution, yielding a value of $\SI{56.7}{dB}$. From the power measured at room temperature ($\SI{-22.7}{dBm}$) and the determined line attenuation ($\SI{74.5}{dB}$), we determine the power of the microwave tone at the transducer input port to be $\SI{-97.2}{dBm}$, leading to a thermal noise equivalent power of $\SI{-153.9}{dBm}$, approximately five orders of magnitude smaller than the instantaneous microwave power used during readout. The thermal noise contributions for other optical powers is straightforwardly determined by taking the calibrated value as a reference and correcting for any measured increase/decrease in thermal noise power by the increased/decreased conversion efficiency. 

\subsection{Shot noise}
\label{sec:added_shot_noise}

The current generated by a photodiode from incident light is given by

\begin{align}
	i = \frac{e \eta_q}{\hbar \omega_o} P_{o},
	\label{shot_noise_current}
\end{align}

with $P_{o}$  the incident optical power, $e$ the elementary charge, $\omega_o$ the optical frequency and $\eta_q$ the quantum efficiency of the detector. The generated electrical noise power is

\begin{align}
	\delta P_e = Z\Var\qty(i) = Z\qty(\frac{e\eta_q}{\hbar \omega_o})^2 \Var\qty(P_o),
	\label{e_noise}
\end{align}

with $Z$ the output impedance of the photodiode. The number of photons $n$ arriving on the detector in a time interval $\Delta t$ obeys Poissonian statistics $\Var\qty(n) = \bar{n}$, which in terms of the incident optical power yields

\begin{align}
	\Var\qty(P_o) = \frac{\hbar \omega_o}{\Delta t} \bar{P}_o.
	\label{P_o_var}
\end{align}

Substituting (\ref{P_o_var}) into (\ref{e_noise}) yields

\begin{align}
	\delta P_e \Delta t = Z\frac{e^2\eta_q^2}{\hbar \omega_o} \bar{P}_o.
\end{align}

For an acquisition window of duration $\Delta t$ we have for the the acquisition bandwidth $\Delta f \approx 1/\pi \Delta t$, such that

\begin{align}
	\delta P_e \approx 2\pi Z\frac{e^2\eta_q^2}{\hbar \omega_o} \bar{P}_o \Delta f,
	\label{e_noise_power}
\end{align}

where we include a factor of two to ensure that the contributions from both the positive and negative frequency components are taken into account. To express equation \ref{e_noise_power} in terms of added noise, we divide by all the amplifications in the signal chain. In our heterodyne scheme, the measured signal is the mixed term of the upconverted transducer microwave input $(P_i)$ with the lower sideband of the LO modulated by a phase EOM $(\alpha P_{LO})$. For a phase EOM at the optimal modulation depth $\beta$, the factor $\alpha$ is given by the square of the maximum value of the first order Bessel function, divided by two (to only consider one sideband) $|J_1(\beta)^2|/2\approx 0.17$~\cite{Yariv2007}. We can then write the electrical signal measured on the spectrum analyzer $P_e$ in function of the transducer microwave input $P_i$

\begin{align}
	\bar{P_e} &= Z(\frac{e\eta_q}{\hbar \omega_o}\bar{P_o})^2 \nonumber \\
	&= Z(\frac{e\eta_q}{\hbar \omega_o}\sqrt{\alpha \bar{P}_{LO}\bar{P_s}})^2 \nonumber \\
	&= Z(\frac{e\eta_q}{\hbar \omega_o})^2 \alpha \bar{P}_{LO} \eta_c \eta_t \frac{\omega_o}{\omega_e}\bar{P_i} 
	\label{e_noise_power_in_function_of_input}
\end{align}

We can now revert equation (\ref{e_noise_power_in_function_of_input}) to find an equation for the transducer microwave input $P_i$ that would lead to the a power $P_e$ measured on the spectrum analyzer

\begin{align}
	P_i = \frac{P_e}{Z\frac{e^2\eta_q^2}{\hbar^2\omega_o \omega_e}\alpha \bar{P}_{LO}\eta_c \eta_t}.
\end{align}

We can now define the added noise $\delta P_{add}$ as the amount of noise on the transducer input that would generate a noise $\delta P_e$ on the spectrum analyzer

\begin{align}
	\delta P_{add} = \frac{\delta P_e}{Z\frac{e^2\eta_q^2}{\hbar^2\omega_o \omega_e}\alpha \bar{P}_{LO}\eta_c \eta_t}.
	\label{P_add}
\end{align}

Substituting (\ref{e_noise_power}) into (\ref{P_add}) then gives the input referred shot noise

\begin{align}
	\delta P_{add} = \frac{2\pi \hbar \omega_e}{\alpha \eta_c \eta_t} \frac{\bar{P}_o}{\bar{P}_{LO}}\Delta f \approx \frac{2\pi \hbar \omega_e \Delta f}{\alpha \eta_c \eta_t}
\end{align}

Where, in the last step, we have used the fact that the power of the local oscillator is much larger than all the other powers in the experiment in a heterodyne scheme. We can now divide by the microwave photon energy $\hbar \omega_e$ in an interval of time $\Delta t = 1/\pi \Delta f$ to get the number of added noise photons on the input of the transducer 

\begin{align}
	n_{add} &\approx  \frac{2}{\alpha \eta_c \eta_t}
	\label{n_add}
\end{align}

\begin{figure}
	\centering
	\includegraphics[width = 1\linewidth]{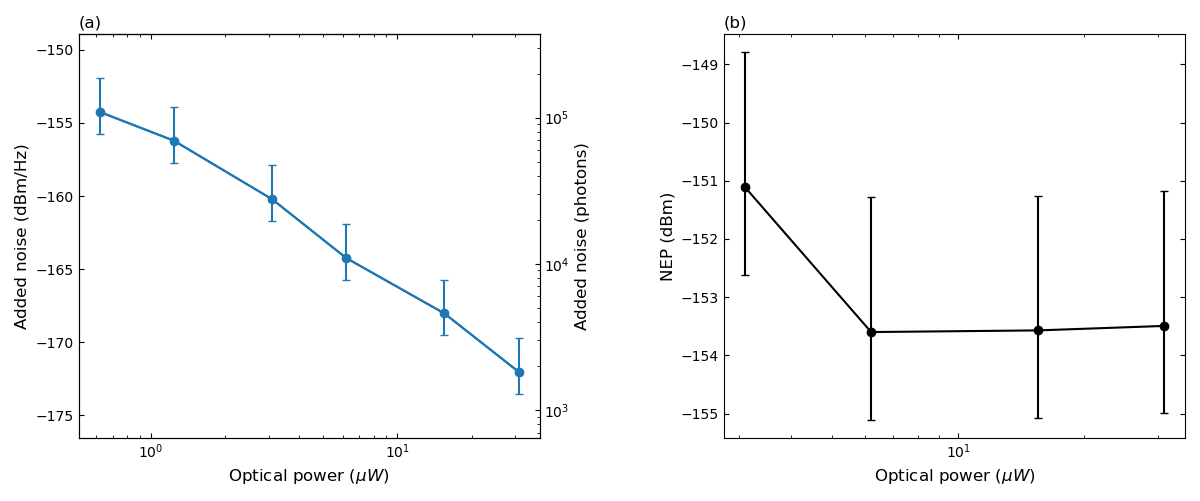}
	\caption{\textbf{Added noise.} (a) Transducer input-referred added noise levels due to heterodyne shot noise in units of $\SI{}{dBm/Hz}$ (left axis) and photons (right axis) as a function of optical pump power. (b) Noise equivalent thermal power as a function of optical pump power.}
	\label{fig:added_noise}
\end{figure}

Fig.~\ref{fig:added_noise} shows the heterodyne shot noise contribution to the added noise as a function of optical power for the device discussed in the main text. According to expectation, the heterodyne contribution decreases with optical power due to the increasing conversion efficiency, reducing from $\sim \SI{1e5}{photons}$ to $\sim \SI{2e3}{photons}$ between $\SI{0.6}{\micro W}$ and $\SI{31}{\micro W}$ optical power. We note that this contribution can be reduced or even fully removed by filtering out the portion of the local oscillator that is not amplifying the signal, increasing $\alpha$ in equation (\ref{n_add}). As discussed previously, we find the added thermal contribution to the added noise to be approximately five orders of magnitude smaller than the instantaneous readout power (see Fig.~\ref{fig:added_noise}). Accordingly, neither of these two contributions can fully explain the noise level observed in the main text. This leaves the possibility of an additional contribution from the pump shot noise, which could introduce uncertainty in the number of photons in the optical cavity, thereby increasing the observed noise. However, further investigation is required to confirm the extent of this effect.  A method for mitigating the pump shot noise is to improve the transducer performance to reach high levels of efficiency at lower levels of optical power.

\section{Pump induced decoherence}
\label{sec:decoherence}

\begin{figure}
	\centering
	\includegraphics[width = 1\linewidth]{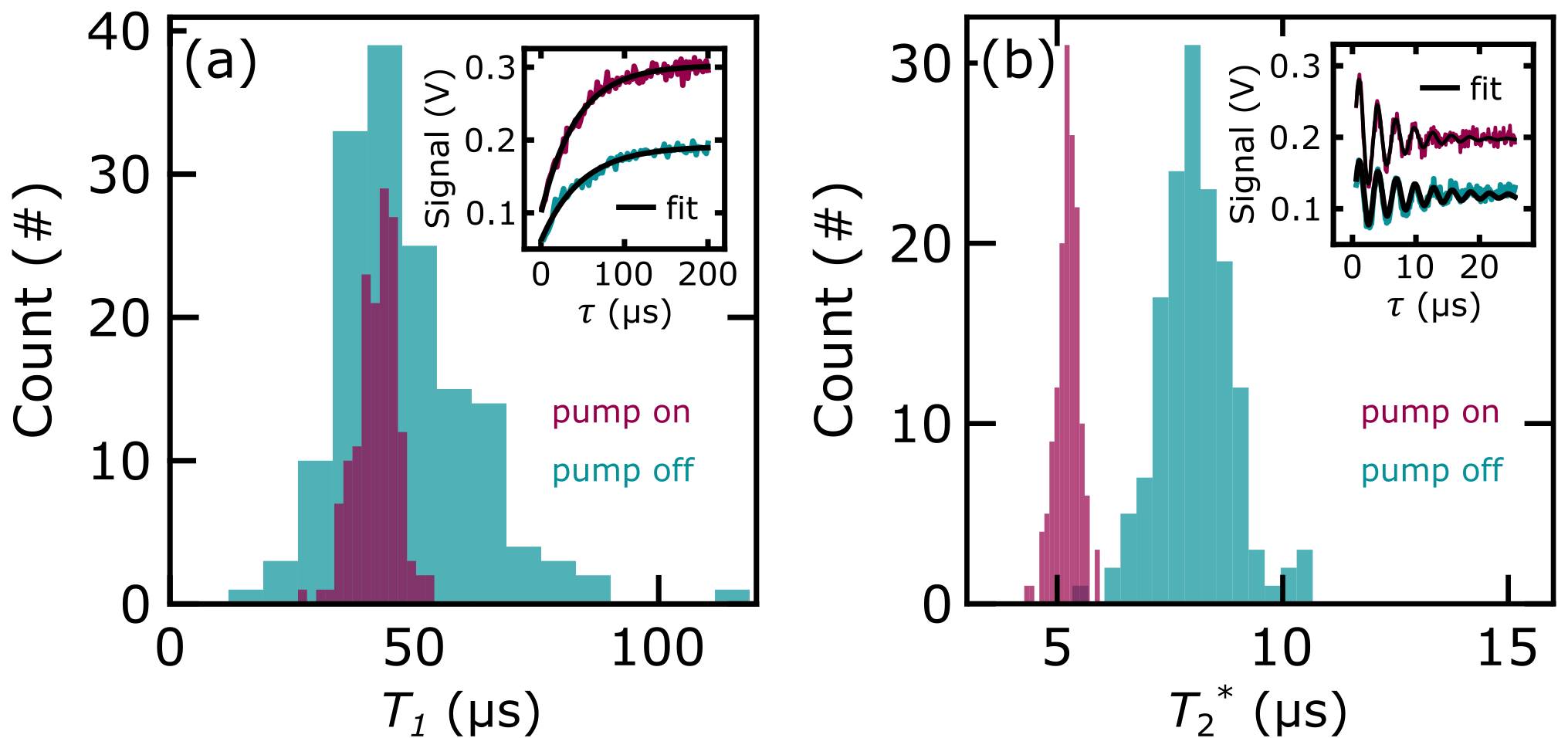}
	\caption{\textbf{Pump-induced decoherence.} In a separate cooldown with a different transducer we repeat the decoherence measurements to determine the effect on the qubit with only one isolator.
		(a) Statistical distributions of 150 $T_1$ measurements, comparing the decay times with the optical pump on (purple) and off (cyan) ($\SI{42.4 \pm 4.4}{\micro s}$ versus $\SI{48.1 \pm 14.1}{\micro s}$, respectively). The inset shows two example $T_1$ decay curves, with fits denoted by the black lines. Similarly, panel (b) compares the statistical distributions of 150 $T_2^*$ measurements with the pump on (purple) and off (cyan) ($\SI{5.2 \pm 0.2}{\micro s}$, versus $\SI{8.1 \pm 0.8}{\micro s}$, respectively), with the inset showing two examples of fitted curves.}
	\label{pumpdecoherence}
\end{figure}

In the main text we demonstrate that the optical pump from the transducer does not cause measurable decoherence of the qubit with three $\SI{20}{dB}$ isolators between the transducer and the qubit. However, the necessity of adding a large number of isolators or filters might strain cryogenic space and thermal budgets, even in multiplexed optical readout scenarios. We therefore also include here data from a separate cooldown with a different transducer and characterize qubit longitudinal and transverse decay times by measuring $T_1$ and $T_2^*$ with only a single $\SI{20}{dB}$ isolator inserted between the transducer and qubit readout resonator. The transducer microwave transmission output is sent to a HEMT microwave amplifier at 4K and used to determine the $T_1$ and $T_2^*$ decay times using microwave-only readout, comparing scenarios with the pump continuously on or off. To obtain a robust estimation, we perform 150 measurements for each scenario. Fig.~\ref{pumpdecoherence}a shows the distribution of the extracted $T_1$ times, with an inset illustrating two example curves and corresponding fits. We find a (statistically insignificant) reduction in mean $T_1$ from (pump off) $\SI{48.1}{\micro s}$ with a standard deviation of $\SI{14.1}{\micro s}$ to (pump on) $\SI{42.4}{\micro s}$ with a standard deviation of $\SI{4.4}{\micro s}$. The difference in signal amplitude between the two scenarios, shown in the inset, can be attributed to a dependence of the transducer microwave resonator transmission on the optical pump power, which exhibits larger transmission with the optical pump on. The increased transmission has previously been shown to originate from the optical field inside the photonic cavity~\cite{weaver2022integrated}. 

We further assess the impact of the pump on the transverse decay rate by measuring $T_2^*$. Fig.~\ref{pumpdecoherence}b shows the statistical distribution of 150 measurements, with the inset illustrating two example traces. We find a statistically significant reduction in $T_2^*$ from (pump off) $\SI{8.1}{\micro s}$ with a standard deviation of $\SI{0.8}{\micro s}$ to (pump on) $\SI{5.2}{\micro s}$ with a standard deviation of $\SI{0.2}{\micro s}$. We note that in the main text we operate the readout with a pulsed optical transducer pump where the optical power is turned off during the delay time of the decay time measurement. With such a pulsed operation we expect the transverse decay time to be longer even with a single isolator.

\begin{figure}
	\centering
	\includegraphics[width=1\linewidth]{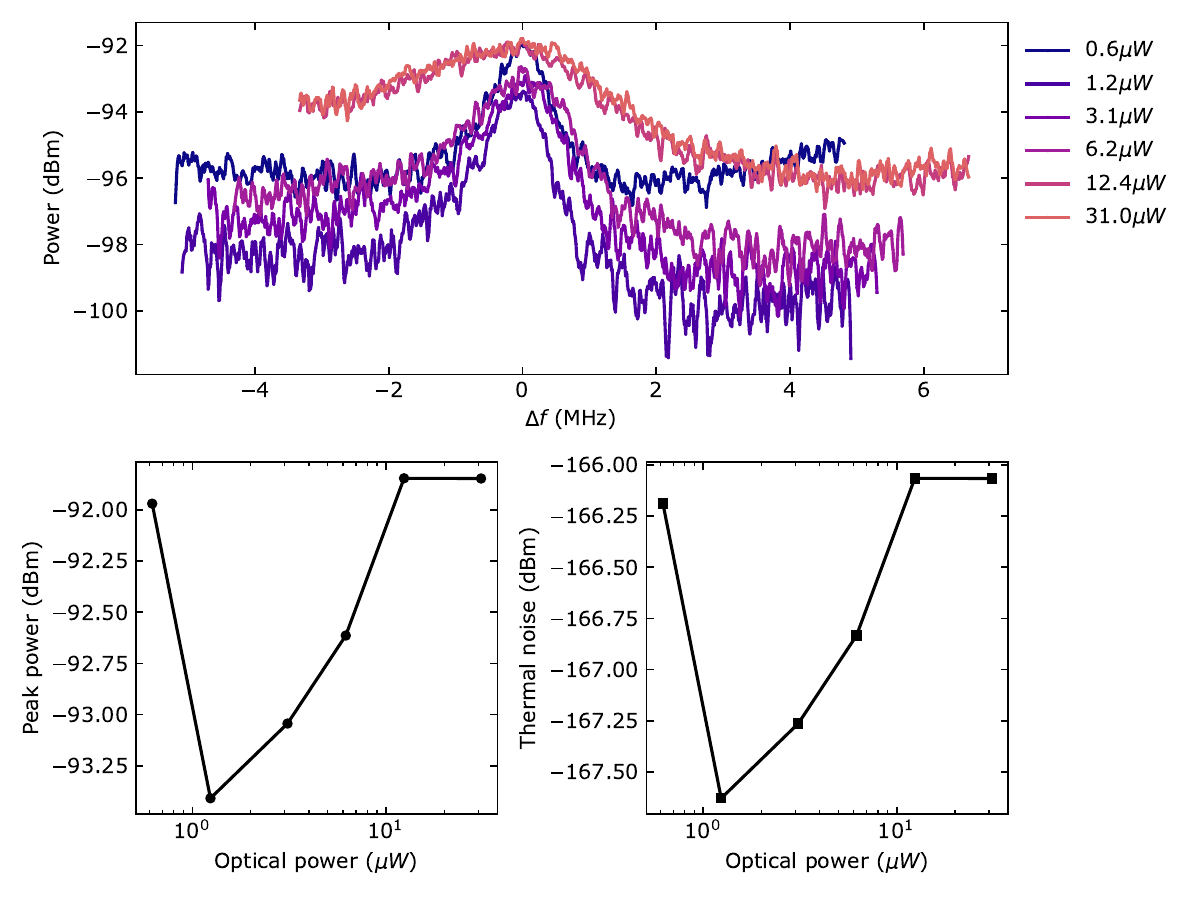}
	\caption{\textbf{Microwave thermal noise.} (a) Thermal microwave emission spectra for different optical powers. (b) Peak thermal microwave power as a function of optical power. (c) Extracted thermal microwave emission at the transducer microwave port.}
	\label{fig:thermal_mw}
\end{figure}

One interesting feature we observe in the data is that the standard deviation of the $T_1$ and $T_2^*$ is decreased with the optical pump on. The reduction in variance could potentially be related to pump-induced thermal excitation of two-level systems in the qubit environment~\cite{schlor2019correlating}. To investigate this, we directly measure the microwave emission from the transducer in the main text. We tune the transducer to its point of peak transduction efficiency and vary the optical pump power. The resulting heating of the mechanical mode leads to thermal microwave photons being emitted via the electromechanical interaction. The resulting microwave signal is subsequently amplified with a travelling-wave parametric amplifier (TWPA), as the first element in the chain and recorded with a spectrum analyser (see Figs.~\ref{fig:thermal_mw}a-b). A microwave calibration tone is simultaneously inserted, which is processed by the same amplifier chain. The cryogenic line attenuation is calibrated using the Stark shift method discussed prior, from which we extract the microwave power on the MXC stage of the dilution refrigerator. The power ratio between the thermal noise and the calibration tone detected at room temperature then identifies the magnitude of the thermal microwave power emitted by the transducer (see Fig.~\ref{fig:thermal_mw}c). As a function of optical power, we find overall a relatively constant thermal emission power of $\sim \SI{-166}{dBm}$, which would be reduced by any isolation placed between the transducer and the readout resonator.

\section{Experimental setup}

\begin{figure}
	\centering
	\includegraphics[width = 1\linewidth]{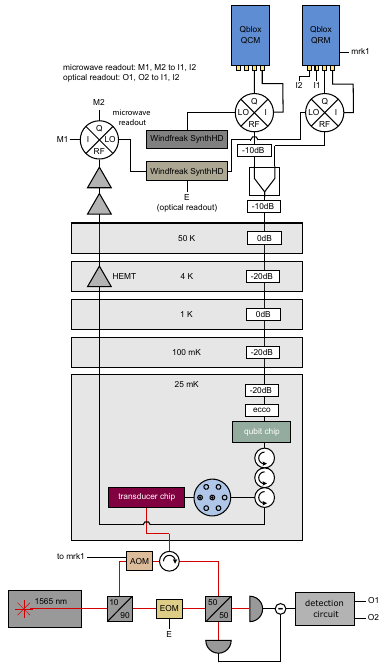}
	\caption{\textbf{Microwave and optical wiring.} Wiring diagram for microwave-only and optical qubit readout.}
	\label{fig:diagram}
\end{figure}

The experimental architecture of microwave-only and optical readout is depicted in Fig.~\ref{fig:diagram}. In-phase and quadrature signals modulated at MHz frequencies are generated by a baseband Qblox Qubit Control Module (QCM) and used to drive the qubit following upconversion to the GHz domain by an I-Q mixer and RF local oscillator. The I-Q mixer is calibrated to reject the upper sideband of the composite microwave signal. A similar procedure is used to generate readout pulses with a Qblox Qubit Readout Module (QRM). We use a power combiner to transmit readout and drive signals towards the single feedline of the qubit chip. Using a cryogenic microwave switch, the signals emerging from the qubit chip can be directed towards either the electrical input port of the transducer (via two $\SI{20}{dB}$ isolators and a circulator) or reflected from the switch towards the HEMT. After reflecting from the transducer microwave port or the switch, the signal is transmitted via a $\SI{20}{dB}$ circulator and an isolator towards a 4K HEMT and two room temperature amplifiers in series. Amplitude and phase information of the signal is then recovered via I-Q down mixing to baseband frequencies and subsequent demodulation by the numerically-controlled oscillator (NCO) of the QRM. For optical readout, the generation of microwave readout signals is identical to the procedure described for microwave-only readout, the only difference being that the RF local oscillator simultaneously drives an EOM in the local oscillator branch of a balanced optical heterodyne detection scheme, which downconverts GHz modulated optical signals to baseband frequencies. An acousto-optic modulator (AOM) is used to pulse the transducer optical pump and turn it on during the readout pulse. Using a custom detection circuit, in-phase and quadrature signals are extracted and demodulated by the NCO of the QRM.

\end{document}